# Weak-Lensing Detection of Intracluster Filaments in the Coma Cluster

Kim HyeongHan[1], M. James Jee[1,2*], Sangjun Cha[1], and Hyejeon Cho[1,3]

[1]Department of Astronomy, Yonsei University, 50 Yonsei-ro, Seoul 03722, Korea.

[2]Department of Physics and Astronomy, University of California, Davis, One Shields Avenue, Davis, CA 95616, USA.

[3]Center for Galaxy Evolution Research, Yonsei University, 50 Yonsei-ro, Seodaemun-gu, Seoul 03722, Republic of Korea.

E-mail: ahrtears54@yonsei.ac.kr, mkjee@yonsei.ac.kr

## Abstract

Our concordance cosmological model predicts that galaxy clusters grow at the intersection of filaments structuring the cosmic web stretching tens of Mega parsecs. Although this hypothesis has been supported by the baryonic components, no observational study has detected the dark matter component of the intracluster filaments (ICFs), the terminal segment of the large-scale cosmic filaments at their conjunction with individual clusters. We report weak-lensing detection of ICFs in the Coma cluster field from the ~12 sq. deg Hyper Suprime-Cam imaging data. The detection is based on two methods: matched-filter technique and shear-peak statistic. The matched-filter technique (shear-peak statistic) yields detection significances of 6.6- (3.1) $\sigma$ and 3.6- (2.8) $\sigma$ for the northern and western ICFs at 110° and 340°, respectively. Both ICFs are highly correlated with the overdensities in the WL mass reconstruction and are well-aligned with the known large-scale (>10 Mpc) cosmic filaments comprising the Coma supercluster.

**Introduction**

Intracluster filament (ICF) is a terminal segment of the cosmic web filament, which, as numerical simulations show[1-3], often penetrates well inside the virial radius of the cluster. We detected ICFs in the Coma cluster ($z = 0.023$; A1656) with the 12 sq. deg high-quality Subaru/Hyper Suprime-Cam (HSC) wide-field imaging data based on the two shear-based approaches (Methods): matched-filter technique[4] (Fig. 1) and shear-peak statistic (Fig. 2). We were able to measure the weak-lensing (WL) signal from the ICFs because 1) the surface mass density of the filament escalates toward the galaxy cluster[1,5], thus increasing its density contrast against the background, and 2) a large number of source galaxies per physical area of the lens boosts the WL statistical power, thanks to the proximity of the Coma cluster (Methods and Supplementary Fig. 1).

**Results**

**Matched-filter statistic**

Figure 1 (see also Supplementary Fig. 2) shows the matched-filter statistic $\Gamma_+$ in polar coordinates together with the reconstructed mass distribution of the Coma cluster. There are three outstanding $\Gamma_+$ peaks whose significance is above 3-$\sigma$ at ~110° (N), ~240° (SE), and ~340° (W). We estimate the significances of the N, SE, and W features to be 6.6-$\sigma$, 4.3-$\sigma$, and 3.6-$\sigma$, respectively, considering both the shot noise and large-scale structure (LSS) along the line-of-sight (LOS). As seen in the mock test (Methods and Supplementary Fig. 3), the cross matched-filter statistic ($\Gamma_\times$) also vanishes at these angles. Among the three ICF candidates, the N and W features are in coincidence with the shear-peak-based detection.

We find that the matched-filter statistic result is consistent with the features identified in the two-dimensional mass reconstruction (Fig. 1). The correlation between the matched-filter statistic peaks and mass overdensities can be clearly seen. The northern and western regions at $r > 1$ Mpc are characterized by overdensities forming radially-aligned linear structures whereas the SE region does not possess such a coherent structure. The linear structure at ~30° is also in good agreement with the location of the matched-filter peak. However, since its matched-filter significance is relatively low (~2.5-$\sigma$), we classify it as an ICF candidate in this paper.

Since there is a strong concentration of mass on the northern ICF (~2.3 Mpc away from the Coma center at ~110°), we investigated whether this mass clump is a main contributor to the detection. When we limit our analysis to the 1 Mpc $< r <$ 2 Mpc annulus excluding this overdense region, we still find the peak at the same orientation with a similar significance (~6-$\sigma$). As demonstrated in our mock test (Supplementary Fig. 3), isotropic mass distributions are not likely to masquerade as significant filament detection in the matched-filter statistic.

One of the extensively studied filaments of the Coma cluster in the literature is the western filament in the direction of the Leo cluster (A1367), which together with A1656 comprises the Coma supercluster[6-11]. The orientation of the western ICF is highly consistent with this filament. It is also remarkable that the central mass substructures within $r < 1$ Mpc form a short (~1 Mpc) filament well-aligned with this ICF.

**Shear-peak count statistic**

Figure 2 shows the shear-peak map and the azimuthal shear-peak density variation. Although it is difficult to detect shear peak density excess from the 2D map alone (left), the shear-peak statistic presents remarkably consistent features with the matched-filter statistic, revealing the density peaks at the same angles. Unlike the matched-filter technique, here we cannot use the same light cone data (Methods) to perform a mock test because of the insufficient resolution. This prevents us from estimating the uncertainty due to the LSS based on the numerical simulation data. Therefore, in order to estimate the statistical significance of the shear-peak statistic, we computed the baseline using the data themselves, assuming that azimuthal average approaches the mean shear peak count in the blank field. Of course, since the Coma cluster is far from a blank field, the baseline obtained in this way is likely to be overestimated. Thus, this approach is a conservative choice. The northern and western peaks possess significances of 3.1-$\sigma$ and 2.8-$\sigma$, respectively. The southeastern peak has a significance of 2.0-$\sigma$, which is much lower than the matched-filter result (4.3-$\sigma$). The feature at ~30° is consistent with the baseline and insignificant in this shear-peak statistic.

**Discussion**

We measured the properties of the ICFs from our Markov Chain Monte Carlo (MCMC) sampling using the parametric model (Eqn. 1 in Method), and the result is presented in Supplementary Table 1. In Figure 3, we display the posterior distributions of the normalization $\kappa_0$ and the characteristic width $h_c$. Although we do not consider the SE peak at $\theta = 240°$ a solid detection, we include it in the analysis to illustrate that its property differs from those of the other two. The northern and western ICFs have a characteristic width of $h_c = 0.29^{+0.05}_{-0.05}$ Mpc and $0.10^{+0.05}_{-0.04}$ Mpc, respectively, and a normalization constant (i.e., the peak density) of $\kappa_0 = 0.0168^{+0.0024}_{-0.0023}$ and $0.0188^{+0.0055}_{-0.0044}$, respectively. On the other hand, in the case of the SE filament candidate, the normalization constant is a factor of two lower ($0.0080^{+0.0018}_{-0.0015}$) than those of the other two. Also, its characteristic width is much larger ($h_c = 1.07^{+0.43}_{-0.30}$ Mpc) than the other two. The properties of the SE feature suggest that perhaps the signal may come from multiple overlapping filaments *if* it is a filament. Assuming a cylindrical symmetry within $4h_c$, the mean densities of the northern and western ICFs are 103 and 115 times the critical density (343 and 383 times the background density) at the Coma redshift, respectively.

In order to compare the ICFs with the larger filaments traced by galaxies around the Coma cluster, we retrieved the galaxy catalog within the $r = 10°$ (~16.8 Mpc) radius of NGC 4874 from the Sloan Digital Sky Survey Data Release 16 (SDSS DR16)[12]. Figure 4 displays the distribution of galaxies around the Coma cluster. We show the directions (magenta arrows) of the four large-scale filaments in the literature[6,7] detected with galaxy overdensities. The orientation of the western ICF is well-aligned with the filament direction in the literature and it is well-traced by the galaxies at similar redshifts to the Coma cluster within this field. The northern ICF orientation is also consistent with the filament direction in the literature. Its presence is not clear in galaxy overdensity within the HSC field ($r < 2°$). Interestingly, the SE peak at ~240°, which we classified as an ICF candidate, is also aligned with the filament suggested by the galaxy overdensity.

The dashed circle in Figure 4 illustrates the relatively small field size of the HSC imaging data. If galaxies are assumed to Poisson-sample the dark matter distribution, it is difficult to trace the ICF with galaxies alone because the region is already crowded with the Coma cluster galaxies, as well as the galaxies belonging to the filaments. In this regard, the matched-filter WL analysis may serve as a useful (and perhaps better) tool to identify the ICFs because the filter is designed to be sensitive to thin linear mass distributions.

A recent study based on the state-of-the-art cosmological hydro-simulation[13] finds that the mean characteristic widths of the short ($L < 9$ Mpc) and long ($L > 20$ Mpc) cosmic filaments are 0.25±0.03 Mpc and 0.24±0.03 Mpc, respectively at $z = 0$. The characteristic width of the northern ICF in Coma approximately corresponds to these mean values, whereas that of the western ICF is on the low end of the distribution. Also, from the result of this study[13], we estimate the mean densities for the short and long filaments within the cylindrical volume at $r <$ $4h_c$ to be 67 and 33 times the background density, respectively. Considering the large filament-to-filament variation[13] and the three-fold filament density increase at the cluster junction[1,5], we find that the properties of the ICFs in Coma are on the high end of the distribution but not exceptional.

The current study is different from the previous studies reporting filament detections in A222/A223[14] and MACSJ0717[15,16] primarily based on weak lensing. The A222/A223 detection was a mass bridge between the two massive clusters, which can be claimed as a long (~18 Mpc) cosmic filament only under the assumption that the filament is aligned along the LOS direction. Currently, there is no evidence supporting such a fortuitous alignment. The MACSJ0717 result is different from the A222/A223 one in that the mass detection extends from the main cluster comprised of two merging clusters to the southeastern X-ray group separated[17] from the main cluster by 1.2 Mpc and then from there further stretches towards the south by 2.5 Mpc. As in the A222/A223 system, under the assumption that this mass extension is aligned along the LOS direction, the structure is interpreted as a projection of the 18 Mpc-long cosmic filament. An X-ray-based filamentary structure detection was reported around A2744[18]. Although these gas structures show signs of correlations with some WL mass and galaxy overdensities, the WL mass reconstruction alone does not convincingly reveal the filamentary structure. Currently, no cosmic web structures extending tens of Mpc, similar to those observed in Coma, have been reported in A2744.

The current discovery of the ICF in Coma has an important implication for the cosmological studies based on cluster mass functions. For the Coma cluster, the sum of the two ICFs may account for up to 10-20% of the total mass within the virial radius, depending on how exactly we define the cluster mass. As in Coma, if a significant fraction of the total mass is generally contained in ICFs within a cluster's virial radius, it becomes imperative to revisit the definition of the cluster mass. In simulations, cluster masses are defined as the total mass within a sphere whose radius is computed for the given density contrast criteria. In observations, typical mass estimation is obtained under the assumption that the density profile follows an analytic description. Given that neither the simulations nor the observations explicitly account for the ICFs, the omission of these structures in both approaches may introduce non-negligible bias in the estimation of cosmological parameters based on cluster masses.

**Methods**

**Feasibility of ICF Detection at Low-$z$**

WL studies of low-redshift ($z \ll 0.1$) galaxy clusters have been considered a difficult task mainly because 1) the lensing efficiency is low and 2) very wide-field imaging data are required. The latter is no longer a limiting factor in today's era, where many wide-field instruments are available. The former is certainly a weakness arising from the geometric effect of the low-redshift lens. However, as we demonstrate below, the weakness is significantly outweighed by the high S/N due to a large number of source galaxies per physical area of the lens. Thus, we argue that the Coma cluster WL provides a considerably better opportunity for low-contrast structure detection than at higher redshift.

We define the WL S/N parameter $\lambda$ as $\lambda = \eta \beta \sqrt{n}/\sigma_{LSS}$, where $\eta$ is the purity of the source population, $n$ is the physical source number density (number of sources per physical area at the lens redshift), $\beta$ is the lensing efficiency, and $\sigma_{LSS}$ is the noise due to the LSS[19] along the LOS direction. The term $\sigma_{LSS}$ decreases with angular scale and thus is smaller at lower redshift because the angular size of the filament is larger. The lensing efficiency is $\beta \equiv \int_{z_l}^{\infty} dz dp(z) D_l D_{ls}/D_s$, where $p(z)$ is the source redshift probability distribution, and $z_l$ is the lens redshift. Supplementary Figure 1 shows that $\beta$ (red) increases with the lens redshift. The Coma lens efficiency is lower nearly by two orders of magnitude than the one at $z_l = 0.5$. However, as the lens redshift decreases, the number of source galaxies per physical area at the lens redshift increases quickly. Therefore, the total lensing S/N per physical area of the lens (blue) in fact increases as the lens redshift decreases, becoming three times higher at the Coma redshift than the S/N value at $z_l = 0.5$. One caveat that one can recall in this S/N analysis is the shear calibration. Since the intrinsic lensing signal is weaker at lower redshift, the requirement for the residual shear calibration error should be more stringent.

**Matched-Filter Statistic**

A matched filter refers to the template (or kernel) that extracts the optimal S/N measurement when it is correlated with the signal. Both the shape of the signal and the noise characteristics are important considerations in designing the filter. In this study, we adopt the formalism presented by Maturi & Merten[4].

Colberg et al.[5] studied the density profiles of filaments from cosmological simulations and found that filaments have well-defined edges characterized by the scale width $h_c$, inside which the density is approximately constant. Outside the edges, the density drops as $h^{-2}$, where $h$ is the perpendicular distance from the "ridge" of the filament. Adopting the result, Maturi & Merten[4]. used the following analytic form to model the filament convergence profile:

$$\kappa(h) = \frac{\kappa_0}{1+(h/h_c)^2}, \tag{1}$$

where $\kappa_0$ is the maximum convergence at the ridge. Note that at the characteristic width $h_c$, the density drops to half the maximum convergence $\kappa_0$. The WL shear $\gamma$ induced by the filament (Eqn. 1) is oriented perpendicular to the filament axis and given by

$$\gamma = \kappa(h). \tag{2}$$

In practice, $\gamma$ (Eqn. 2) is obtained from the ellipticities of source galaxies. Assuming that our coordinate origin lies at one end of the filament and $\theta_f$ is the orientation angle of the filament measured counterclockwise from the x-axis, $\gamma$ is related to the galaxy ellipticity components $\epsilon_1$ and $\epsilon_2$ via $\gamma = \epsilon_1 \cos[2(\theta_f - \pi/2)] + \epsilon_2 \sin[2(\theta_f - \pi/2)]$. If we let $\gamma_i$ be the estimate of $\gamma$ from the $i$-th source galaxy, our matched-filter statistic is given by

$$\Gamma = \frac{1}{W}\Sigma_i \gamma_i \Psi(x_i, y_i), \tag{3}$$

where $\Psi(x_i, y_i)$ is the filter/weight at the position $(x_i, y_i)$ of the $i$-th source galaxy and $W = \Sigma_i \Psi(x_i, y_i)$. The filter $\Psi$ is obtained by considering the expected shear signal of the filament and the noise due to the LSS. In Fourier space, the optimal filter is given as follows:

$$\widehat{\Psi}(\mathbf{k}) = \hat{\tau}(\mathbf{k})/P_n(k), \tag{4}$$

where $\hat{\tau}(\mathbf{k})$ is the Fourier transform of the expected shear from the filament (Eqn. 2) and $P_n(k)$ is the noise power spectrum including the LSS and intrinsic galaxy shape noise. Note that since we let the estimator (Eqn. 3) include the proper normalization (i.e., $1/W$), we omit a normalization constant in Equation (4).

In WL, the signal comes from the scalar potential and thus produces only "E-mode" signals. The "B-mode" signal is obtained by rotating galaxy ellipticities by 45° and should be consistent with zero if there is no significant residual systematics. In a similar fashion, we can define the corresponding terms in the matched-filter statistic $\Gamma$. We use $\Gamma_{+(\times)}$ to denote the E(B)-mode statistic and refer it to the tangential (cross) component hereafter. Note that the cross component $\Gamma_\times$ does not necessarily become zero even in the absence of the systematics because any coherent alignment of galaxies tilted 45° with respect to the filter orientation can produce non-zero values.

Maturi & Merten[4] provides the following analytic form for the variance of $\Gamma$ (Eqn. 3):

$$\sigma_\Gamma^2(\theta) = \frac{1}{2W^2}\Sigma_i |\gamma_i|^2 \Psi^2(x_i, y_i). \tag{5}$$

We find that Equation (5) does not properly include the LSS contribution and leads to an overestimation of the filament detection significance. Thus, in this study, we use the publicly available kappaTNG light cone (convergence) data (http://columbialensing.org)[20] constructed from the cosmological hydrodynamic simulations IllustrisTNG300-1[21-26] to properly estimate the statistical significance of our filament detection. We select the convergence dataset for the source redshift at $z_s = 0.5$ since it is the closest match to the effective source redshift ($z_{eff} \sim 0.7$) estimated for the current Coma cluster WL data. From the 100 5 × 5 deg² convergence fields, we generated 900 subfields matching the current Coma field size (~12 sq. deg). Then, we produced mock shear catalogs at the positions of the Coma source galaxies. Finally, we applied the matched filter to the shear catalogs and estimated the variance at each angle due to the LSS effect. We added the two uncertainties from this LSS contribution and the shot noise (Eqn. 5) in quadrature to compute the total uncertainty. Supplementary Figure 2 displays the matched filter statistic and its uncertainty derived from the Coma field.

In order to evaluate the performance of the matched-filter technique for filament detection, we created mock shear catalogs and applied the method. Supplementary Figure 3 shows two examples of the mock test results. In the top panel, we assumed that a Coma-like cluster ($M_{200c} = 8 \times 10^{14}\ M_\odot$) at $z = 0.02$ is placed at the intersection of two filaments at 10° and 70° with a linear mass density of $m_f = 1.2 \times 10^{14}\ M_\odot$ Mpc$^{-1}$. These mock simulation setup parameters are deliberately chosen to resemble the properties of the Coma cluster and its filaments. We find that the matched filter statistic successfully identifies the two filaments and reaches maxima at the correct orientations. The cross-shear statistic ($\Gamma_\times$) possesses much lower amplitudes and vanishes at the orientation of the filament. In the bottom panel, the setup is identical to the one in the top panel except that another Coma-like cluster at the same redshift is placed at $\theta = 135°$ about 1.4 Mpc away from the field center. This is to test whether the presence of massive halos can masquerade as filaments in our matched-filter statistic. The result shows that although a small bump is detected at $\theta = 135°$, its amplitude is substantially lower than the peaks due to the filaments, which illustrates that the matched filter is much more sensitive to the low-density anisotropic linear structure than to the high-density isotropic mass peak. In the two examples shown in

Supplementary Figure 3, we did not include the noise due to the intrinsic galaxy shape dispersion and LSS for illustrative purposes. The significance of the bump due to the second galaxy cluster drops below our detection threshold when we include the due noise.

**Shear-Peak Statistic**

In addition to the matched-filter statistic, another useful method for ICF detection proposed by Maturi & Merten[4] is the shear-peak statistic. In general, shear peaks refer to projected over-dense regions in WL mass map and their statistic as a function of significance is a powerful probe of cosmological parameters[27-32]. High S/N peaks are found at the locations of individual massive clusters whereas true low S/N peaks are mostly produced by the projection of many low-mass halos along the LOS direction, rather than by single low-mass halos. Because of their low S/N nature, a significant fraction of low S/N shear peaks is false positives. Since the presence of ICFs boosts the probability of low S/N shear peak detection, we can trace the ICF with their counting statistic.

In this study, we employ an aperture mass filter derived from a truncated NFW profile to minimize the influence from the LOS LSS. The convergence of the projected NFW profile[33] is $K_{\mathrm{NFW}}(x) = 1/(1+x)^2$, where $x = \vartheta/\vartheta_{\mathrm{s}}$, $\vartheta$ is the angular separation, and $\vartheta_{\mathrm{s}}$ is the scale radius. Its tangential shear profile is approximated as $G_{\mathrm{NFW}}(x) = 2\ln(1+x)/x^2 - 2/x(1+x) - 1/(1+x)^2$, and at the outskirts we apply the Gaussian truncation function: $G(x) = G_{\mathrm{NFW}}(x)\exp(-\vartheta^2/2\vartheta_{\mathrm{out}}^2)$, where $\vartheta_{\mathrm{out}}$ is the truncation radius.

The mass map produced by this aperture filter was treated as if it was an astronomical image of a crowded stellar field. We detected the shear peaks using Sextractor[34] (https://github.com/astromatic/sextractor) and measured the number of peaks by azimuthally scanning a pan-shape region at every degree. The pan-shape region has an opening angle of 30 degrees and covers a radial extent from 1 Mpc to 2.8 Mpc. We manually set the background level to zero in our SExtrator run.

**Data Reduction**

We used Subaru HSC observations in $g$ and $r$ bands. HSC is an 870-Mega pixel (116 2k × 4k CCDs) mosaic CCD camera with a pixel scale ~0.16″ and a ~1.5° diameter field of view. The observations were conducted with seven different pointings, which together forms a hexagonal footprint whose largest dimension exceeds 4.5°. The radius of the largest complete circle within the mosaic field is ~1.9°. The details of the observations are given in Supplementary Table 2.

Single-frame processing (overscan, bias, flat, dark, etc.) was performed with the HSC Pipeline[35], which was built on the prototype pipeline developed for the Rubin Telescope Legacy Survey of Space Time (LSST). The pipeline also includes robust calibration of the World Coordinate System (WCS), which stores the astrometric and distortion solutions in the SIP format. Since our WL pipeline uses the SWARP package, which only understands the TPV format, we converted the astrometric header of each frame using the sip-to-pv (https://github.com/stargaser/sip_tpv.git) script[36]. These calibrated files were combined with SWARP to create giant (97k pixel × 91k pixel) mosaic images. We generated two versions of mosaic images. The first version is produced using all available frames except for defective ones. This product provides deep photometry and is also useful for object detection. The second mosaic is created with the frames whose seeing is better than 0.85″. This image is optimized for galaxy shape measurement. Supplementary Fig. 4 displays the color-composite image created with these large mosaic images, whose total effective area is ~12 sq. deg.

**Object Detection and Photometry**

We ran SExtractor for the object detection and photometry. We used the dual-image mode to obtain object colors with consistent segmentation information. We provided SExtractor with a weight image created by SWARP for the detection and a root mean square error image for the photometry. For object detection, we required objects to satisfy the criteria DETECT_THRESH = 2 and DETECT_MINAREA = 5. We set the deblending parameters with the blending threshold (DEBLEND_NTHRESH) of 64 and the minimal contrast (DEBLEND_MINCONT) of $10^{-4}$ to identify the overlapping objects. We used MAG_ISO for color estimation and MAG_AUTO for luminosity measurement. We performed photometric calibration using the SDSS DR16[12].

**Source Selection**

We select source galaxies as follows. First, we choose objects between $23 < r < 26$. Because of the proximity of the Coma cluster, most of the faint galaxies in this selection are likely to be background sources. We do not impose a color cut, since the member galaxy contamination is estimated to be negligible in this magnitude range. To confirm this, we compare the number density distributions between the Coma and COSMOS2020[37] fields (Supplementary Fig. 5). In the latter, there is no Coma-like cluster. Therefore, if our magnitude-based source selection includes non-negligible fraction of the Coma members, the Coma field must show an excess in the selection window. The excellent agreement between the two fields verifies that indeed the cluster galaxy contamination can be ignored within the source selection window.

The objects with a half-light radius ($r_h$) smaller than ~0.5″ are discarded to avoid stellar contamination. Our forward-model shape measurement utilizes the MPFIT[38] module, which reports the status and stability of the shape fitting and the ellipticity uncertainty. We only keep the sources with the stable fitting (STATUS = 1) and a small ellipticity uncertainty (< 0.4).

Some spurious sources cannot be identified with the method described above. They include false detections around bright stars forming concentric circles and unidentified cosmic rays near the field edges. As for the former, we deselect the objects by applying large circular masks on the stars brighter than $r = 13$ mag. The spurious objects near the edges of the six flanking fields are not included in our final source catalog because we reject the sources whose distances are $r > 2.8$ Mpc from the Coma cluster center. We note that there are no spurious objects on the edges of the six flanking fields facing toward the cluster center because the regions overlap the central and other flanking fields. The number density in the final source catalog is ~38 arcmin$^{-2}$. Since it is impossible to derive photometric redshifts only from the two filters, we utilize the publicly available GOODS-S photometric redshift catalog[39] to estimate the effective redshift of the source population. After applying the same source selection criteria to the GOODS-S photometric redshift catalog and taking into account the difference in image depth, we determine the effective source redshift to be $z_{eff} = 0.72$. Because of the proximity of the Coma cluster, the critical surface density is not sensitive to the source redshift. For example, even if we assume $z_{eff} = 1$, the critical surface density changes by approximately 1%.

**Shear Measurement**

We measured galaxy shapes from the mosaic image. This requires us to model the point-spread-function (PSF) for each CCD frame per exposure and to carefully propagate the PSF model at the location of the galaxy on the coadd. The PSF model for each CCD frame was constructed through Principal Component Analysis (PCA) based on the observed stars and polynomial interpolation[40,41]. At each galaxy location, we first identify all contributing frames and then stack their PSF models after applying due rotations. This PSF modeling scheme enables us to capture the sophisticated PSF variation patterns across the mosaic image including sudden discontinuities at the CCD boundaries. The PSF residual (observed star ellipticity – model PSF ellipticity) correlation function shows that the correlation amplitude is below $10^{-6}$ on angular scales greater than 1′ (left panel of Supplementary Fig. 6).

Once we obtain a secure PSF model at the location of a source galaxy, we convolve an elliptical Gaussian function with the PSF and fit the resulting profile to the object[42-46]. Since the PSF-convolved elliptical Gaussian profile differs from the observed galaxy profile, the so-called model bias occurs. In addition, the ellipticity measurement based on the maximum likelihood is biased because of the nonlinear relation between pixel noise and shear, often termed "noise bias." Moreover, a significant fraction of galaxies is affected by blending effects[47]. In order to address the mixture of these sources of shear measurement bias, we use image simulations and compare input (true) shears with output (measured) shears. The simulation shows that the additive factor, which mainly arises from an imperfect PSF model, is negligible. We find that the star-galaxy ellipticity cross correlation is at the $10^{-6}$ level on angular scales greater than 10′ (right panel of Supplementary Fig. 6). Both the PSF residual and star-galaxy correlation diagnostics verify that the residual WL systematic error is negligible. The WL pipeline we employed here participated in the GRavitational lEnsing Accuracy Testing 3 (GREAT3) challenge[48] and was publicly validated to be one of the best-performing methods in this blind test.

**Data Availability:** The raw Subaru/HSC imaging data used for the current study are publicly available. The processed mosaic images and data points in the article figures are available on request from the authors.

**Code Availability:** Our custom data processing codes are available on request from the authors.

## Acknowledgements

M. J. J. acknowledges support for the current research from the National Research Foundation (NRF) of Korea under the programs 2022R1A2C1003130 and RS-2023-00219959.

## Author contributions

K. H. reduced the data, measured weak-lensing signals and filaments, and wrote the manuscript. M. J. J. designed the research, created the reduction pipeline, coded the filament detection filter, and wrote the manuscript. S. C. provided the mass reconstruction based on convolutional neural network. H. C. identified the filaments based on the Coma galaxy distribution.

## Competing interests

N/A

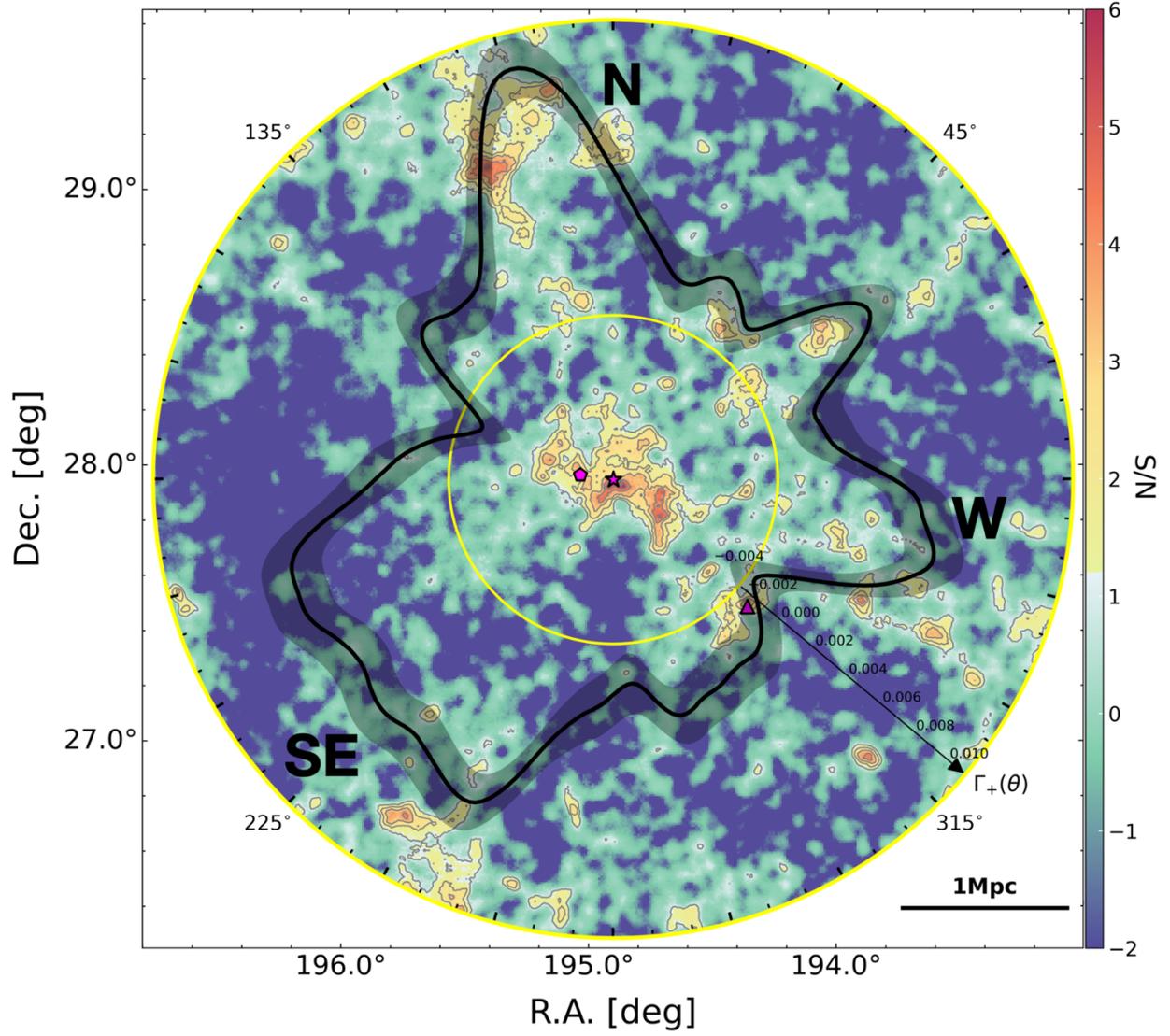

**Figure 1: Mass reconstruction and matched filter statistic of the Coma cluster.** For mass reconstruction, we used the convolutional neural network (CNN) method[49] after training the deep learning with a wide-field (3.5 deg × 3.5 deg) convergence field. We verify that the mass reconstruction result is consistent with the version generated with the conventional method[50]. Magenta star, cross, and triangle represent NGC 4874, NGC 4889, and NGC 4839, respectively. The inner and outer yellow solid circles represent the radii ($r = 1$ and 2.8 Mpc, respectively) of the annulus, where we detected the ICF signal. The black solid curve and shade illustrate the matched-filter statistic ($\Gamma_+$) and its uncertainty (shot noise + LSS effect), respectively, in polar

coordinates (see also Supplementary Fig. 2). The matched-filter statistic correlates well with the mass overdensity. Throughout the paper, we assume a flat $\Lambda$CDM cosmology characterized by $h = 0.7$ and $\Omega_{m,0} = 1 - \Omega_{\Lambda,0} = 0.3$.

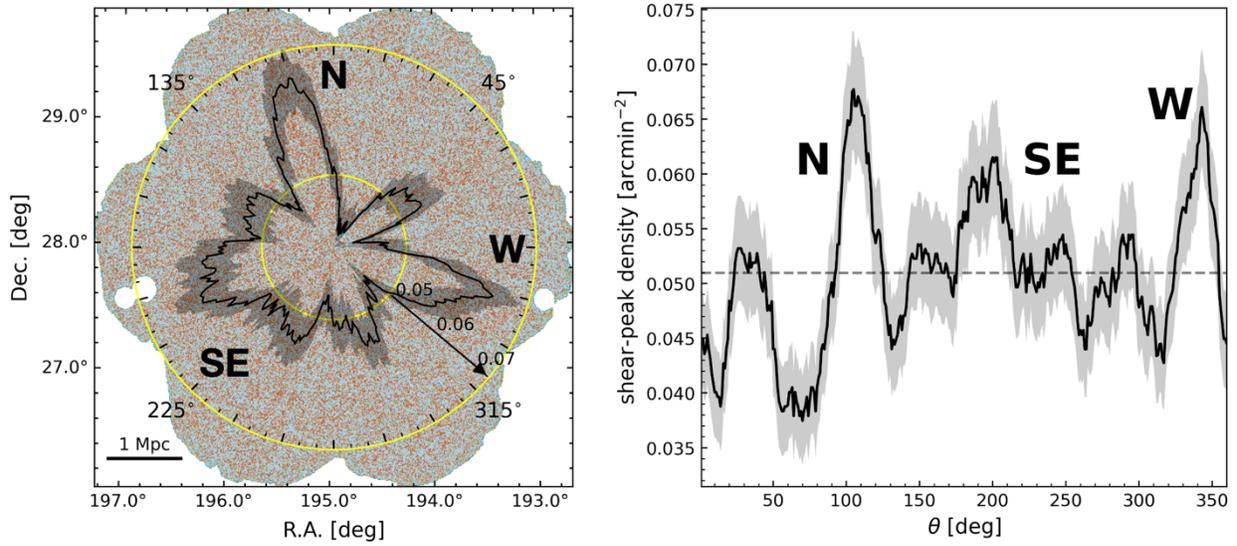

**Figure 2: Intracluster filament detection based on shear-peak statistic. Left**: We display the WL shear-peak map created with the $5 \times 10^{12}$ $M_\odot$ NFW halo filter. The inner and outer yellow solid circles represent $r = 1$ and 2.8 Mpc, respectively. The black solid curve and shade indicate the shear-peak density and its uncertainty in polar coordinates. Circular blank regions are the stellar masks. **Right**: Shear-peak statistic as a function of angle. Black solid line represents the density of shear peaks in sq. arcmin and shade 1-$\sigma$ uncertainty from the Poisson statistics. Dashed horizontal line is the azimuthal mean. The northern and western peaks are in excellent agreement with the matched-filter result (Fig. 1).

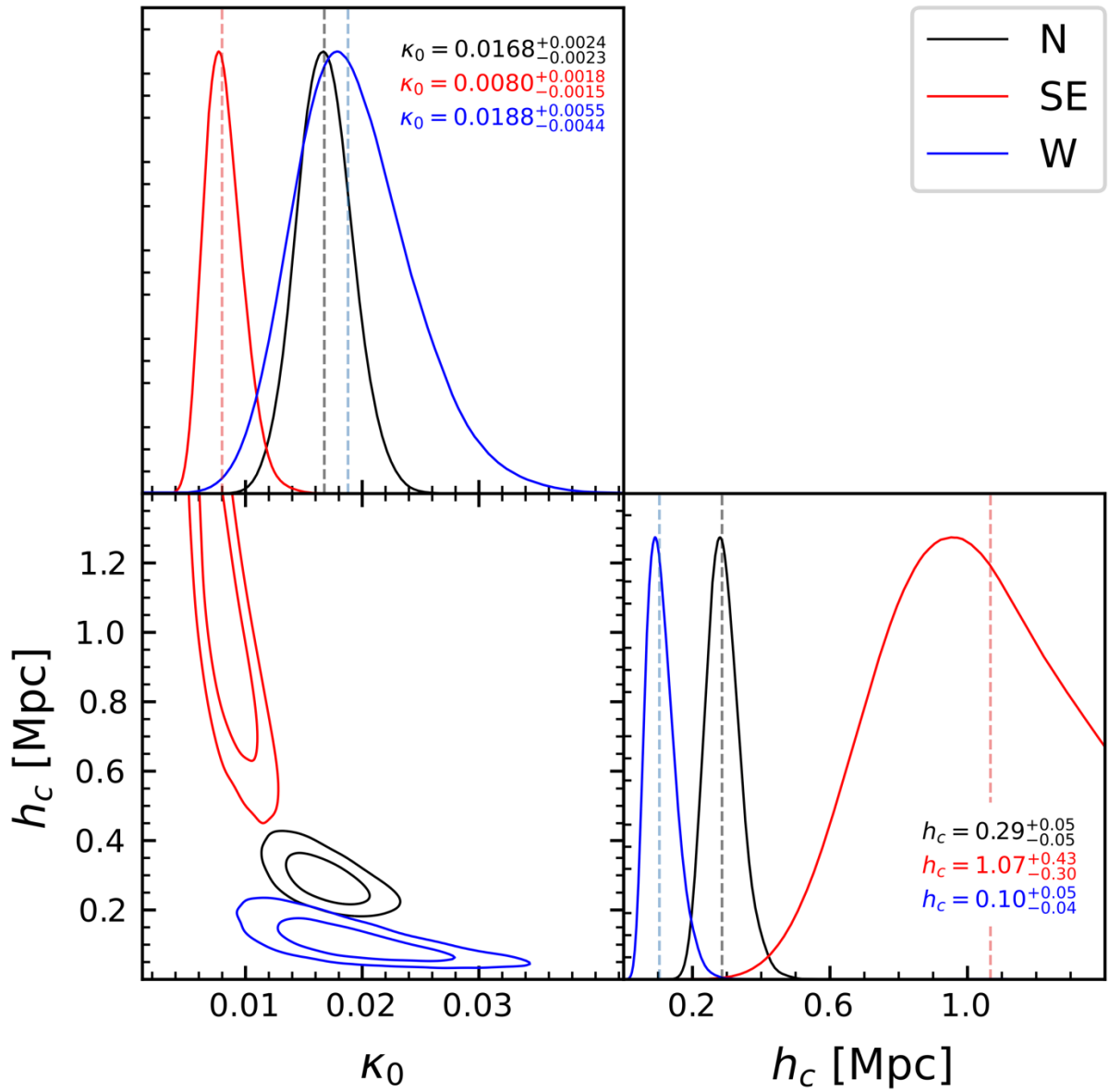

**Figure 3: Posterior distributions of the filament model parameters estimated from the Markov Chain Monte Carlo sampling.** Inner and outer contours represent the 68% and 95% confidence limits, respectively. Dashed lines indicate the median. See text for interpretation.

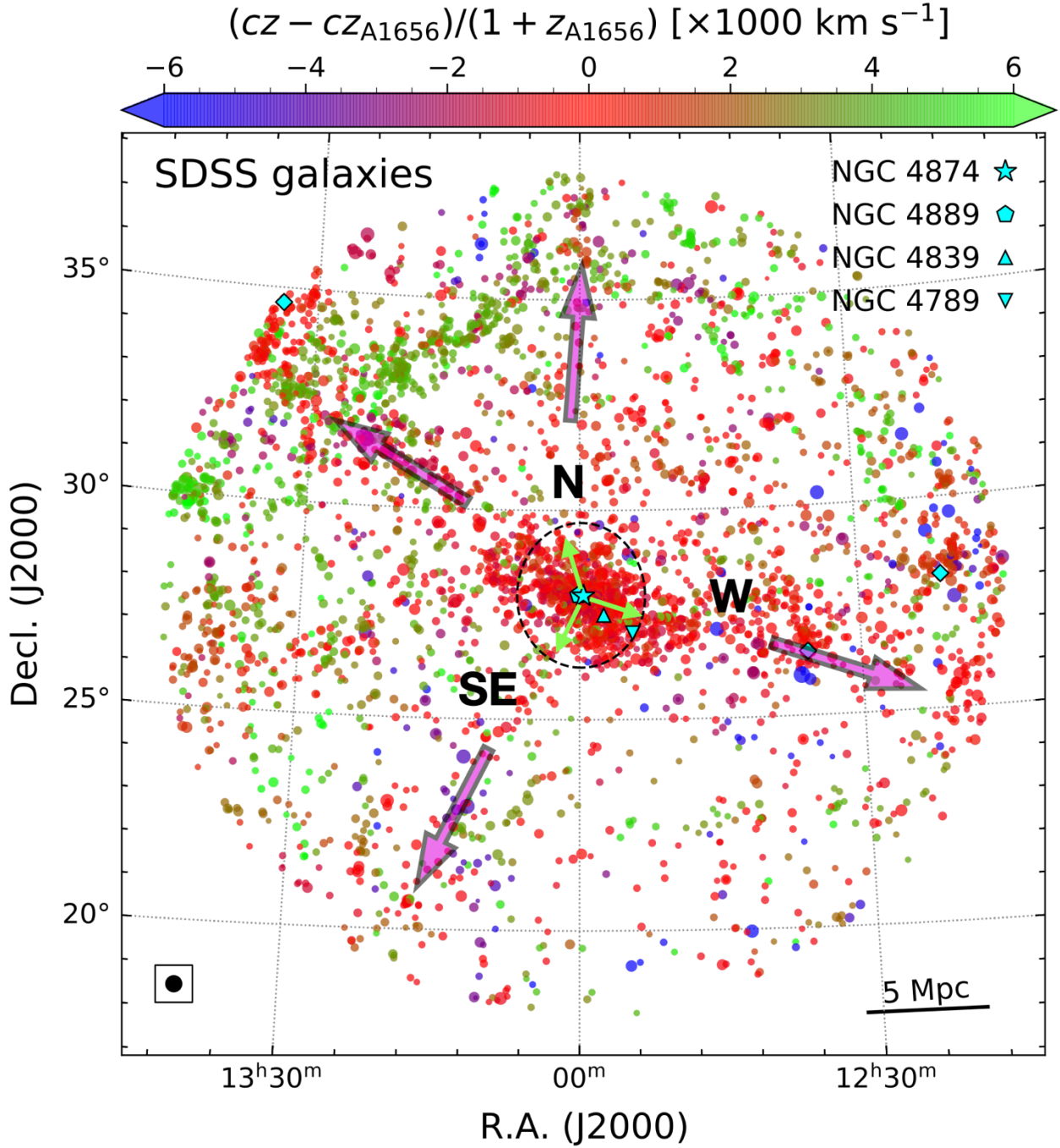

**Figure 4: Alignment of the ICFs with the large-scale filaments.** The LOS velocity from the SDSS DR16 catalog[12] is color-coded. The symbol size is proportional to the SDSS-*r* band magnitude. The size of the filled black circle in the lower left corner represents the brightness of NGC 4874. Neighboring galaxy groups are marked with cyan diamonds. The black dashed *r* =

2.8 Mpc circle indicates the HSC coverage, where we performed WL analysis. The light green arrows within it show the orientations of the ICFs detected in the current study based on WL signals. Magenta arrows indicate the approximate cosmic-web filament directions detected with the galaxy distributions[6,7]. The WL-based ICFs are well-aligned with the galaxy-based large-scale filaments.

**References**

1. Cautun, M., van de Weygaert, R., Jones, B. J. T. & Frenk, C. S. Evolution of the cosmic web. *Mon. Not. R. Astron. Soc.* **441**, 2923-2973 (2014).
2. Kuchner, U. et al. Mapping and characterization of cosmic filaments in galaxy cluster outskirts: strategies and forecasts for observations from simulations. *Mon. Not. R. Astron. Soc.* **494**, 5473-5491 (2020).
3. Rost, A. et al. The ThreeHundred: the structure and properties of cosmic filaments in the outskirts of galaxy clusters. *Mon. Not. R. Astron. Soc.* **502**, 714-727 (2021).
4. Maturi, M., & Merten, J. Weak-lensing detection of intracluster filaments with ground-based data. *Astron. Astrophys.* **559**, A112 (2013).
5. Colberg, J. M., Krughoff, K. S., & Connolly, A. J. Intercluster filaments in a $\Lambda$CDM Universe, *Mon. Not. R. Astron. Soc.* **359**, 272-282 (2005).
6. Mahajan, S., Singh, A., & Shobhana, D. Ultraviolet and optical view of galaxies in the Coma supercluster. *Mon. Not. R. Astron. Soc.* **487**, 4336-4347 (2018).
7. Malavasi, N. et al. Like a spider in its web: a study of the large-scale structure around the Coma cluster. *Astron. Astrophys.* **634**, A30 (2020).
8. Williams, B. A., & Kerr, F. J. The distribution of the spiral galaxies in the direction of the Coma/A 1367 supercluster. *Astron. J.* **86**, 953-980 (1981).
9. Fontanelli, P. The Coma/A1367 filament of galaxies. *Astron. Astrophys.* **138**, 85-92 (1984).
10. Kim, K. T., Kronberg, P. P., Giovannini, G. & Venturi, T. Discovery of intergalactic radio emission in the Coma-A1367 supercluster. *Nature* **341**, 720-723 (1989).
11. Gavazzi, G., Carrasco, L., & Galli, R. The 3-D structure of the Coma-A 1367 supercluster: Optical spectroscopy of 102 galaxies. *Astron. Astrophys.* **136**, 227-235 (1996).
12. Ahumada, R. et al. The 16[th] Data Release of the Sloan Digital Sky Surveys: First Release from the APOGEE-2 Southern Survey and Full Release of eBOSS Spectra. *Astrophys. J. Suppl.* **249**, 3
13. Galárrage-Espinosa, D., Langer, M., & Aghanim, D. et al. Relative distribution of dark matter, gas, and stars around cosmic filaments in the IllustrisTNG simulation. *Astron. Astrophys.* **661**, A115 (2022).
14. Dietrich, J. et al. A filament of dark matter between two clusters of galaxies. *Nature* **487**, 202-204 (2012).
15. Jauzac, M. et al. A weak lensing mass reconstruction of the large-scale filament feeding the massive galaxy cluster MACS J0717.5+3745. *Mon. Not. R. Astron. Soc.* **426**, 3369-3384 (2012).
16. Jauzac, M. et al. Growing a 'cosmic beast': observations and simulations of MACS J0717.5+3745. *Mon. Not. R. Astron. Soc.* **481**, 2901-2917 (2018)
17. Ma, C., Ebeling, H., & Barrett, E. An X-ray/optical study of the complex dynamics of the core of the massive intermediate-redshift cluster MACS J0717.5+3745. *Astrophys. J.* **693**, L56 (2009).
18. Eckert, D. et al. Warm-hot baryons comprise 5-10 per cent of filaments in the cosmic web. *Nature* **528**, 105-107 (2015).
19. Hoekstra, H., How well can we determine cluster mass profiles from weak lensing? *Mon. Not. R. Astron. Soc*. **339,** 1155-1162 (2003).
20. Osato, K., Liu, J., & Haiman, Z. $\kappa$TNG: effect of baryonic processes on weak lensing with IllustrisTNG simulations. *Mon. Not. R. Astron. Soc*. **502,** 5593-5602 (2021).

21. Nelson, D. et al. The IllustrisTNG simulations: public data release. *Comput. Astrophys. Cosmol.* **6**, 2 (2019).
22. Pillepich, A. et al. First results from the IllustrisTNG simulations: the stellar mass content of groups and clusters of galaxies. *Mon. Not. R. Astron. Soc*. **475,** 648-675 (2018).
23. Springel, V. et al. First results from the IllustrisTNG simulations: matter and galaxy clustering. *Mon. Not. R. Astron. Soc*. **475,** 676-698 (2018).
24. Nelson, D. et al. First results from the IllustrisTNG simulations: the galaxy colour bimodality. *Mon. Not. R. Astron. Soc*. **475,** 624-647 (2018).
25. Naiman, J. P. et al. First results from the IllustrisTNG simulations: a tale of two elements - chemical evolution of magnesium and europium. *Mon. Not. R. Astron. Soc*. **477,** 1206-1224 (2018).
26. Marinacci, F. et al. First results from the IllustrisTNG simulations: radio haloes and magnetic fields. *Mon. Not. R. Astron. Soc*. **480,** 5113-5139 (2018).
27. Reblinsky, K., Kruse, G., Jain, B., & Schneider, P. Cosmic shear and halo abundances: analytical versus numerical results. *Astron. Astrophys.* **351**, 815-826 (1999).
28. Dietrich, J. P., & Hartlap, J. Cosmology with the shear-peak statistics. *Mon. Not. R. Astron. Soc*. **402,** 1049-1058 (2010).
29. Kratochvil, J. M., Haiman, Z., & May, M. Probing cosmology with weak lensing peak counts. *Phys. Rev. D* **81**, 043519 (2010).
30. Maturi, M., Angrick, C., Pace, F., & Bartelmann, M. An analytic approach to number counts of weak-lensing peak detections. *Astron. Astrophys*. **519**, A23 (2010).
31. Bard, D. et al. Effect of Measurement Errors on Predicted Cosmological Constraints from Shear Peak Statistics with Large Synoptic Survey Telescope. *Astrophys. J*. **774**. 49 (2013).
32. Marian, L., Smith, R. E., Hilbert, S., & Schneider, P. The cosmological information of shear peaks: beyond the abundance. *Mon. Not. R. Astron. Soc*. **432,** 1338-1350 (2013).
33. White, M., & Kochanek, C. S. Constraints on the Long-Range Properties of Gravity from Weak Gravitational Lensing. *Astrophys. J*. **560**. 539-543 (2001).
34. Bertin, E., & Arnouts, S. SExtractor: Software for source extraction. *Astron. Astrophys*. **117**, 393-404 (1996).
35. Bosch, J. et al. The Hyper Suprime-Cam software pipeline. *Publ. Astron. Soc. Jap* **70**, S5 (2018).
36. Shupe, D. L. et al. More flexibility in representing geometric distortion in astronomical images. *Software and Cyberinfrastructure for Astronomy II.* Proc. SPIE **8451***,* 84511M (2012).
37. Weaver, J. R. et al. COSMOS2020: A Panchromatic View of the Universe to z ~ 10 from Two Complementary Catalogs. *Astrophys. J. Suppl.* **258**, 11 (2022).
38. Markwardt, C. B. Non-linear Least-Squares Fitting in IDL with MPFIT. *Astron. Soc. Pacif. Conf. Ser.,* **411**, 251-254 (2009).
39. Dahlen, T. et al. A Detailed Study of Photometric Redshifts for GOODS-South Galaxies. *Astrophys. J*. **724**. 425-447 (2010).
40. Jee, M. J. et al. Principal Component Analysis of the Time- and Position-dependent Point-Spread Function of the Advanced Camera for Surveys. *Publ. Astron. Soc. Pac.* **199**, 1403 (2007).
41. Jee, M. J., & Tyson, J. A. Toward Precision LSST Weak-Lensing Measurement. I. Impacts of Atmospheric Turbulence and Optical Aberration. *Publ. Astron. Soc. Pac.* **123**, 596 (2011).
42. Finner, K. et al. MC$^2$: Subaru and Hubble Space Telescope Weak-lensing Analysis of the Double Radio Relic Galaxy Cluster PLCK G287.0+32.9. *Astrophys. J.* **851**, 46 (2017).

43. HyeongHan, K. et al. Discovery of a Radio Relic in the Massive Merging Cluster SPT-CL J2023-5535 from the ASKAP-EMU Pilot Survey. *Astrophys. J.* **900**, 127 (2020).
44. Finner, K. et al. Exemplary Merging Clusters: Weak-lensing and X-Ray Analysis of the Double Radio Relic, Merging Galaxy Clusters MACS J1752.0+4440 and ZWCL 1856.8+6616. *Astrophys. J.* **918**, 72 (2021).
45. Cho, H. et al. Multiwavelength Analysis of A1240, the Double Radio-relic Merging Galaxy Cluster Embedded in an ~80 Mpc-long Cosmic Filament. *Astrophys. J.* **925**, 68 (2022).
46. Finner, K. et al. Hubble Space Telescope and Hyper-Suprime-Cam Weak-lensing Study of the Equal-mass Dissociative Merger CIZA J0107.7+5408. *Astrophys. J.* **942**, 23 (2023).
47. Dawson, W. A. et al. The Ellipticity Distribution of Ambiguously Blended Objects. *Astrophys. J.* **816**. 11 (2016).
48. Mandelbaum, R. et al. GREAT3 results - I. Systematic errors in shear estimation and the impact of real galaxy morphology. *Mon. Not. R. Astron. Soc*. **450,** 2963-3007 (2015).
49. Hong, S. E. et al. Weak-lensing Mass Reconstruction of Galaxy Clusters with a Convolutional Neural Network. *Astrophys. J.* **923**, 266 (2021).
50. Kaiser, N., & Squires, G. Mapping the Dark Matter with Weak Gravitational Lensing. *Astrophys. J.* **404**, 441 (1993).

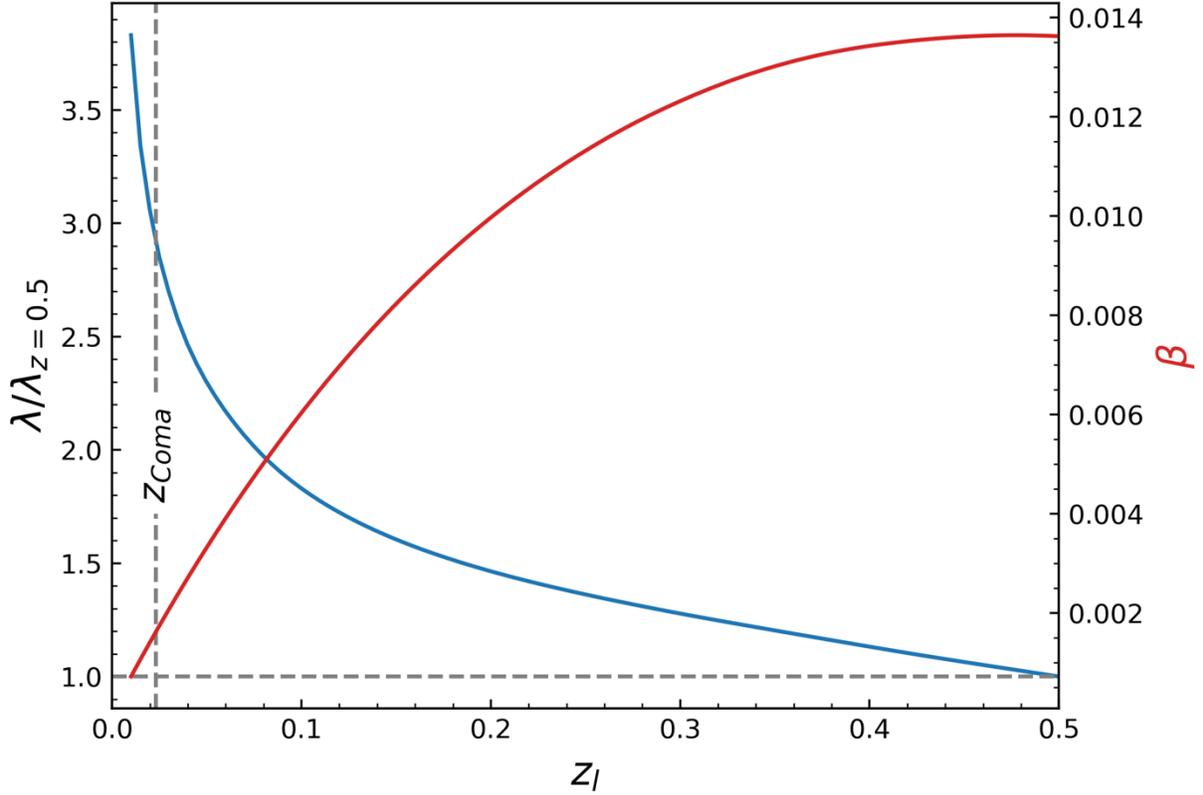

**Supplementary Figure 1: WL S/N per physical area variation with the lens redshift.** Red solid line is the lensing efficiency $\beta$. Blue solid line is the WL S/N per physical area ($\lambda$) normalized by the value at $z = 0.5$. While the lensing efficiency is reduced approximately two orders of magnitude as the lens redshift decreases from $z = 0.5$ to the Coma redshift ($z = 0.023$), the net WL S/N per physical area at the lens redshift increases approximately by a factor of three. This net gain is due to the combined effect of the higher score density $n$ and lower LSS noise $\sigma_{LSS}$ per physical area at the Coma redshift. Here we assume that the purity $\eta$ is the same across the lens redshift. In fact, $\eta$ is also expected to be higher at the Coma redshift because of the proximity.

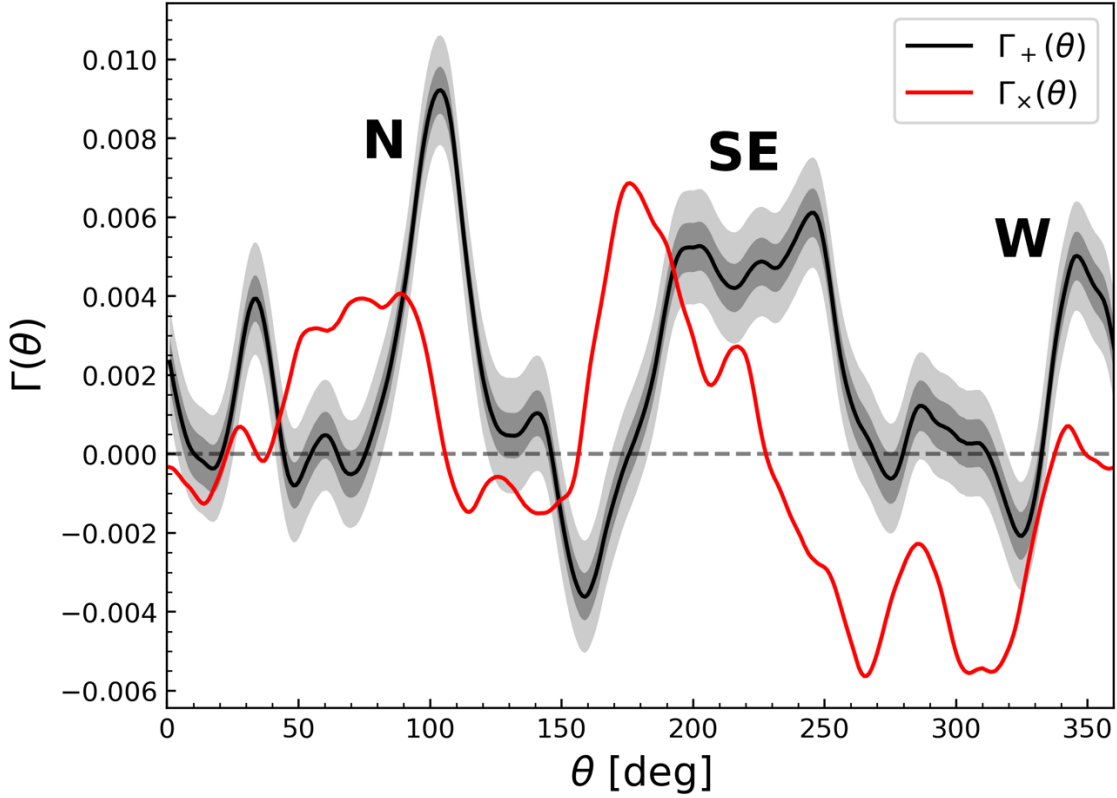

**Supplementary Figure 2: Matched-filter detection of the ICFs in the Coma cluster.** Black (red) solid line represents the tangential (cross)-shear component. Dark shade indicates the shot noise (Eqn. 5). Light shade includes both the shot noise and the noise due to the LSS effect. Three ICFs (N, SE, and W) are detected with a significance level higher than $3\sigma$. The cross-component crosses zero at these locations.

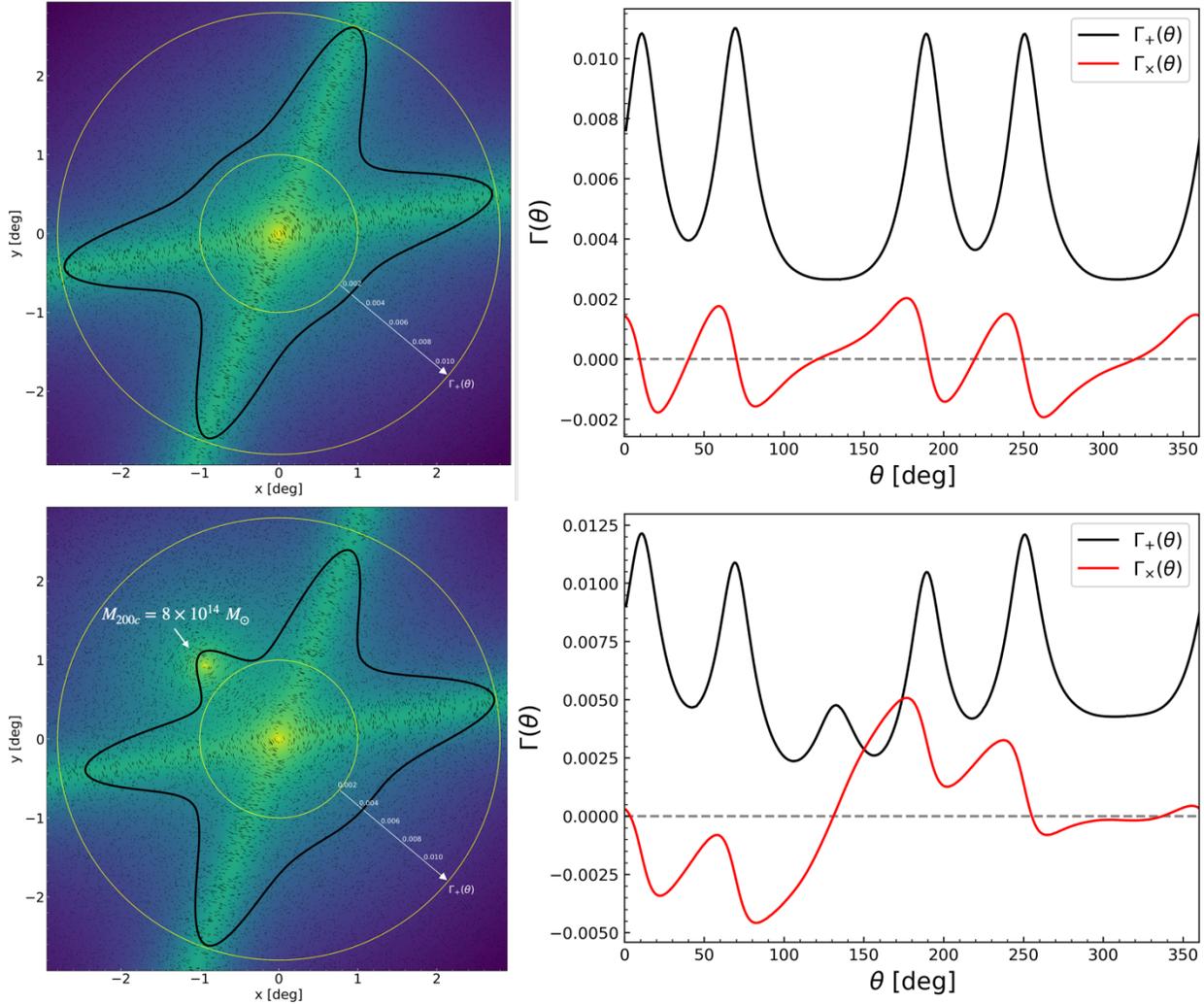

**Supplementary Figure 3: Matched-filter detection with a mock filament-cluster configuration. Top Left:** Mock convergence map with two filaments and an NFW halo displayed in log scale. Mock filaments are oriented at 10° and 70° with an NFW halo at the center. The linear mass density of the filament $m_f = 1.2 \times 10^{14}\ M_\odot$ Mpc$^{-1}$ and the halo mass $M_{200c} = 8 \times 10^{14}\ M_\odot$ are motivated by our measurements in Coma. Whiskers show the resulting WL shear. Inner and outer yellow circles indicate $r = 1$ and 2.8 Mpc, respectively. **Top Right:** Black (red) solid line indicates the tangential (cross)-component $\Gamma_+$ ($\Gamma_\times$). The tangential-shear component is also shown in the left panel (black solid) in polar coordinates. **Bottom:** Same as the top panel except that another $M_{200c} = 8 \times 10^{14}\,M_\odot$ halo is placed (white arrow) at $\theta = 120°$, ~1.2 Mpc apart from the center. The tangential component $\Gamma_+$ is sensitive to the filamentary mass structure and peaks at the correct angles. Although the off-center halo gives rise to a small peak at $\theta = 120°$, its height is much weaker than those arising from the real filamentary structures. Here we show the version that does not include intrinsic shape noise for illustration purposes. With the inclusion of the shape noise, the tangential-component peak due to the halo is below our detection threshold.

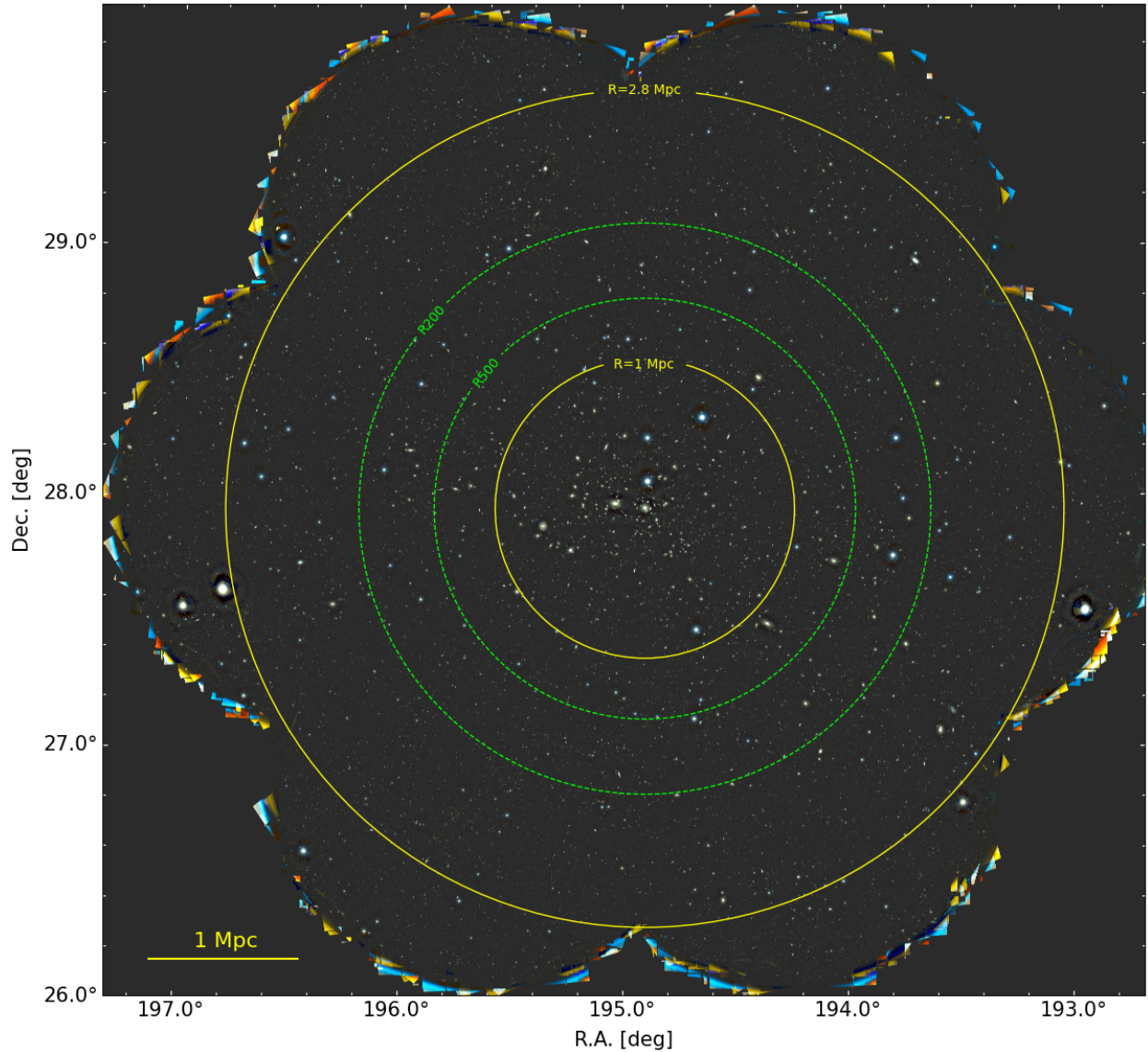

**Supplementary Figure 4: Color-composite image of the Coma cluster created with 7 pointings of Subaru/HSC observations.** Blue, green, and red represent the intensities in *g*, *g+r*, and *r* filters, respectively. The $R_{500}$ and $R_{200}$ radii defined by our weak-lensing mass estimate ($M_{200c} = 8 \times 10^{14} \, M_\odot$) are indicated by green dashed circles. Yellow circles denote the annulus (1 - 2.8 Mpc) we use to detect the ICFs. The total effective area is ~12 sq. deg.

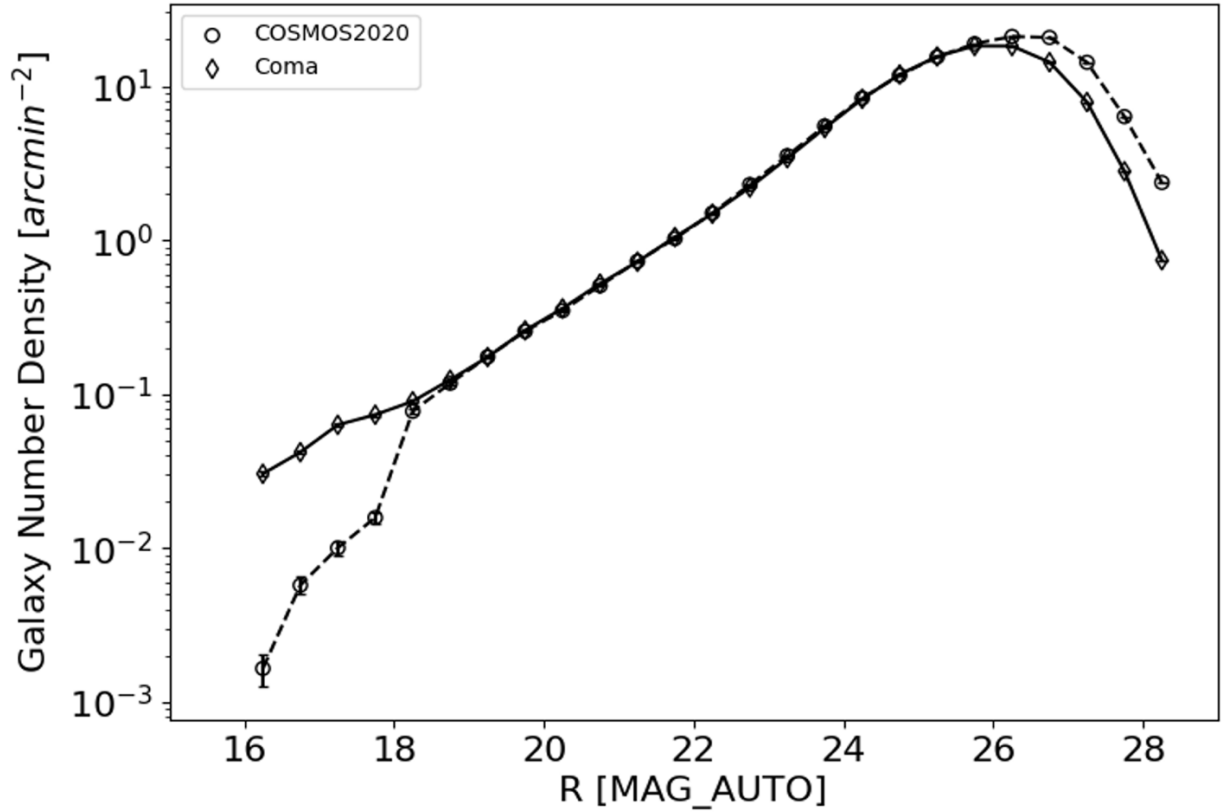

**Supplementary Figure 5: Comparison of the number densities between the Coma and COSMOS2020 fields.** In the bright magnitude regime ($R < 19$), the Coma field clearly shows an excess due to the presence of the cluster members. Within the source selection window ($23 < R < 26$), the number densities between the two fields are in excellent agreement, which illustrates that the cluster member contamination is negligible. Since the COSMOS2020 field is deeper, its number density is higher in the faint magnitude regime ($R > 26$). The error bars are drawn using the Poisson statistics.

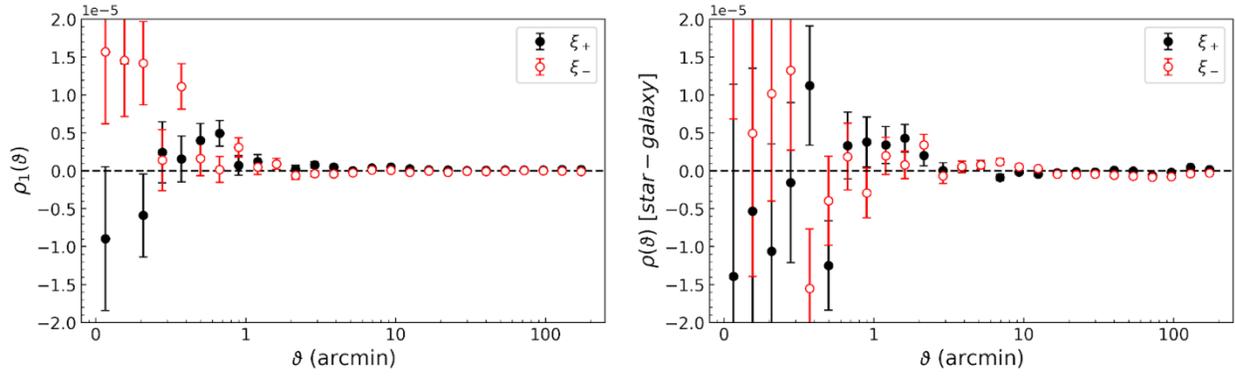

**Supplementary Figure 6: PSF and shape diagnostics. Left**: The auto-correlation of the residual ellipticity (observed star ellipticity – model PSF ellipticity). The amplitude is ~$10^{-7}$ level on angular scales greater than 10′. **Right**: The star-galaxy ellipticity cross-correlation. Again, the small correlation here shows that the shear systematic error due to the imperfect PSF model is negligible. The error bars represent $1\sigma$ uncertainties.

**Supplementary Table 1:** Physical Properties of the Intracluster Filaments

| Filament | Orientation (deg) | Peak Significance (1) ($\sigma$) | $\kappa_0$ | $h_c$ (Mpc) | Linear Mass Density (2) ($\times 10^{14}\ M_\odot\ \text{Mpc}^{-1}$) |
|---|---|---|---|---|---|
| N | 110 | 6.6 (15.5) (3) | $0.0168^{+0.0024}_{-0.0023}$ | $0.29^{+0.05}_{-0.05}$ | $1.36^{+0.31}_{-0.30}$ |
| SE | 240 | 4.3 (9.9) (3) | $0.0080^{+0.0018}_{-0.0015}$ | $1.07^{+0.43}_{-0.30}$ | $2.41^{+1.11}_{-0.82}$ |
| W | 340 | 3.6 (8.1) (3) | $0.0188^{+0.0055}_{-0.0044}$ | $0.10^{+0.05}_{-0.04}$ | $0.55^{+0.31}_{-0.25}$ |

(1) Based on the matched-filter technique. (2) Estimated within the width of $2h_c$. (3) The numbers in parenthesis are the significance when we include only the shot noise.

**Supplementary Table 2: Summary of HSC observations**

| Filter | Exposure Time (s) | Observation Date | ⟨FWHM⟩ (arcsec) | Limiting Magnitude (1) |
|---|---|---|---|---|
| HSC-$g$ | 7,350 | 2016, 2017 | 1.3 | 23.2 |
| HSC-$r$ (2) | 2,880 | 2016 | 1.4 | 24.6 |
| HSC-$r$ (3) | 2,160 | 2016 | 0.7 | 25.0 |

(1) We quote the $5\sigma$ limiting magnitude of a point source. (2) Used for object detection. (3) Used for shape measurement.